\documentclass[a4paper,11pt]{article}
\usepackage{jinstpub} 
\usepackage{hyperref}
\usepackage{graphicx}

\title{\boldmath Performance testing of gas-tight portable RPC for muography applications}

\author[a,1]{V. Kumar\note{Corresponding author.},}
\author[a]{S. Basnet,}
\author[a]{E. Cortina Gil,}
\author[a]{P. Demin,}
\author[a]{R. M. I. D. Gamage,}
\author[a]{A. Giammanco,}
\author[b]{R. Karnam,}
\author[a]{M. Moussawi,}
\author[c]{A. Samalan,}
\author[c,d]{M. Tytgat,}
\author[e]{A. Youssef}

\affiliation[a]{Centre for Cosmology, Particle Physics and Phenomenology (CP3), Universit\'e catholique de Louvain, Louvain-la-Neuve, Belgium}
\affiliation[b]{Centre for Medical and Radiation Physics (CMRP), National Institute of Science Education and Research (NISER) Bhubaneswar, India}
\affiliation[c]{Department of Physics and Astronomy, Ghent University, Ghent, Belgium}
\affiliation[d]{Physics Department, Vrije Universiteit Brussel, Brussels, Belgium}
\affiliation[e]{Multi-Disciplinary Physics Laboratory, Optics and Fiber Optics Group, Faculty of Sciences, Lebanese University, Hadath, Lebanon}

\emailAdd{vishalkmrjswl@gmail.com}

\abstract{This paper reports the latest developmental efforts for a position-sensitive glass-based Resistive Plate Chamber (RPC) and a multi-channel Data AcQuisition (DAQ) system tailored for muon tracking in muography applications. The designed setup prioritizes portability, aiming for field applications where both the detector and the DAQ operate effectively in external environmental conditions. Comprehensive discussions on hardware development activities and signal processing techniques are included, incorporating noise filtering to enhance the accurate detection of real muons. A muon absorption measurement has also been carried out to understand the behavior of these detectors from an application perspective. }


\begin{document}
\maketitle
\flushbottom

\section{Introduction}
\label{sec:intro}
Muography is a technique used for scanning by analysis of muon interactions with a target object~\cite{Bonechi:2019ckl,IAEA2022}. The interactions occur through various mechanisms, with absorption and multiple coulomb interaction being the most prominent ones. Important applications of muography include anomaly detection for maritime security~\cite{barnes2023cosmic}, monitoring of volcanic activity~\cite{muraves2022}, scanning of monuments and artefacts to safeguard our heritage~\cite{Moussawi:2023zkd}, etc. Muography provides non-destructive sub-surface imaging techniques with high penetration power. 
The accuracy of muography scans depend on several factors, with some of the important parameters being the tracking system geometry, the efficiency and position resolution of the muon tracking detector system. Differently from most particle physics experiments, however, particle detectors for muography must often fulfill additional requirements, such as portability, autonomy and robustness, in order to be potentially deployed in logistically challenging locations.

To address these factors, our project proposes the use of a gas-tight portable RPC detector for muon detection, offering a cost-effective solution with high efficiency and position resolution that can be scaled to a large area~\cite{MuographyBook-RPCchapter}. Our portable Resistive Plate Chamber (RPC) detector prototypes are being developed for non-destructive inspection of relatively small volumes, such as archaeological artefacts, where conventional radiography methods cannot be employed. 

This paper is organized as follows:
Section~\ref{sec:Setup} describes the setup used in the performance measurements reported in this paper, while Section~\ref{sec:daq} illustrates the components of our Data AcQuisition (DAQ) system and Section~\ref{sec:analysis} reports our current offline selection and the performance results obtained. In Section~\ref{sec:Absorption}, we show the outcome of a measurement of the absorption of muons in a lead block, which illustrates the capability of our current prototypes for muography applications. 
Finally, in Section~\ref{sec:conclusions} we summarize this work and outline some of the next steps in this project.

\section{Experimental Setup}
\label{sec:Setup}

The experiments reported in this paper were conducted using a gas-tight, portable, glass-based RPC specifically designed for muography applications, as detailed in~\cite{samalan2023small, gamage2022portable}. The RPC comprises an aluminum chamber housing two graphite-coated glass plates measuring $20 \times 20~{\rm cm}^2$, positioned with a 1 mm gas gap between them. The detector operates with a commonly used RPC gas mixture consisting of 95.2$\%$ Freon, 0.3$\%$ SF6, and 4.5$\%$ isobutane.

\begin{figure}[htbp]
\centering
\includegraphics[width=0.95\textwidth, trim=0cm 0 0cm 0cm, clip]{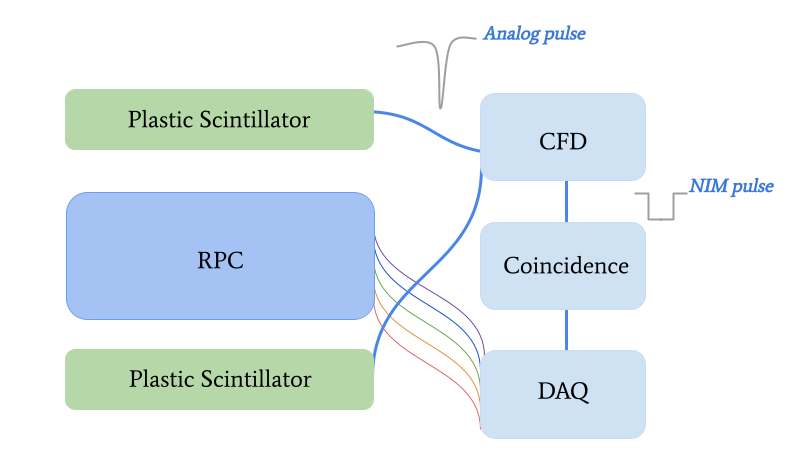}
\caption{Experimental setup used for threshold scan and efficiency measurement.\label{fig:Exp_setup}}
\end{figure}

The readout board, constructed from PCB material, features 16 strips with a strip width of 0.9 cm and a pitch of 1.0 cm, yielding therefore an active surface of $16 \times 16~{\rm cm}^2$. 
Each strip on the readout board incorporates a $110~\Omega$ impedance resistor connected to the ground before linking to a flat cable that extends towards the DAQ system (described in Section~\ref{sec:daq}). 
Figure \ref{fig:Exp_setup} illustrates a schematic diagram of the experimental setup designed for conducting threshold and voltage scans. These scans are intended for both noise reduction and efficiency measurement. In this setup, the RPC is positioned between the two plastic scintillators, whose coincidence signal serves as both a trigger for the DAQ system and a reference for the efficiency measurement. The signals from the plastic scintillators undergo digitization through Constant Fraction Discriminator (CFD) before being fed into the DAQ.

\section{Data Acquisition system}
\label{sec:daq}
The DAQ system consists of three integrated components, each serving a specific function within the experimental setup. These modules are designed for providing high voltage to the RPC, gathering signals from the readout strips through a 34-pin ribbon cable (comprising 16 signal pins and 18 ground pins), and processing these signals to generate comprehensible data for subsequent analysis. The three primary modules of the DAQ system are outlined below:

\subsection{Front-End Electronics}
The front-end electronics employed in this project were originally designed for the large-area RPCs employed in the CMS experiment~\cite{abbrescia2000new} at CERN. This system features a front-end ASIC chip based on 0.8 $\mu m$ BiCMOS technology, offering a charge dynamic range spanning from 20 fC to 20 pC. Each ASIC incorporates eight identical channels, comprising an amplifier, zero-crossing discriminator, monostabilizer, and a differential line driver. The present system has 4 such ASIC chips providing signal processing capacity for 64 channels.

\subsection{FPGA + CPU module}

The Trenz Electronic TE0720-03-1CF (FPGA + CPU) module is built around the Xilinx Zynq XC7Z020 system-on-chip (SOC)~\cite{xilinx_docs}, seamlessly integrating a dual-core ARM Cortex-A9 CPU and an FPGA on a singular integrated circuit. With 75 differential LVDS input/output interfaces having a 5 ns clock cycle, the module establishes connections to the Front-End Electronics. It operates autonomously, serving as a self-contained unit capable of hosting FPGA configuration files and data acquisition software. Facilitating external communication, the module is equipped with Ethernet and USB interfaces.

\subsection{High Voltage supply}
The detector plates are biased using the DPS Mini Series high voltage module, specifically designed for high precision and stability. This module is chosen for its low ripple and noise characteristics. Its compact design makes it advantageous, delivering precise high voltage control through the onboard CPU mentioned in the previous section.

\section{Data Analysis}
\label{sec:analysis}

Before delving into the intricacies of data analysis, it is crucial to comprehend the data structure, which is recorded with a FPGA clock frequency of 200 MHz, ensuring a time accuracy of 5 ns. A time adjustment is made by delaying the RPC signals by 30 ns, to take into account that the RPC signal precedes the trigger signal from the plastic scintillators, which is delayed by the photo-multiplier electron transit time of approximately 30 ns. The digital signal samples are continuously recorded within a time window commencing 620~ns before and concluding 620~ns after the trigger signal. It is noteworthy that certain results presented herein are derived from cross-checks performed with refined and optimized analyses. These analyses were conducted post-conference, underscoring the developmental nature of the detector prototype.

\begin{figure}[htbp]
\centering
\includegraphics[width=.82\textwidth, trim=0cm 0 0cm 0cm, clip]{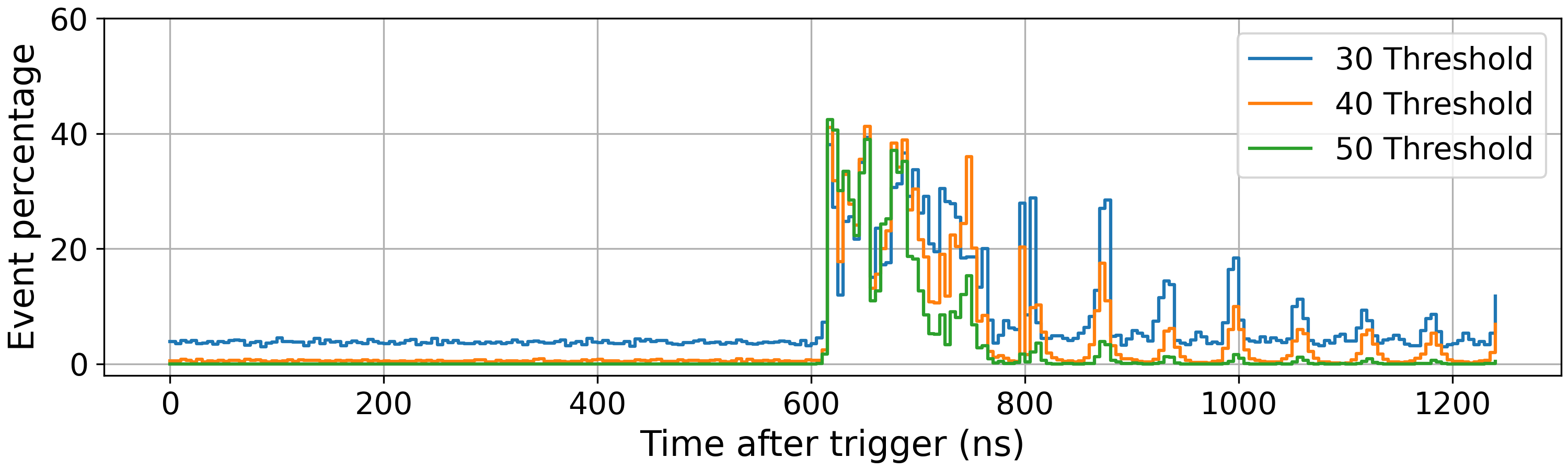}
\includegraphics[width=.82\textwidth, trim=0cm 0 0cm 0cm, clip]{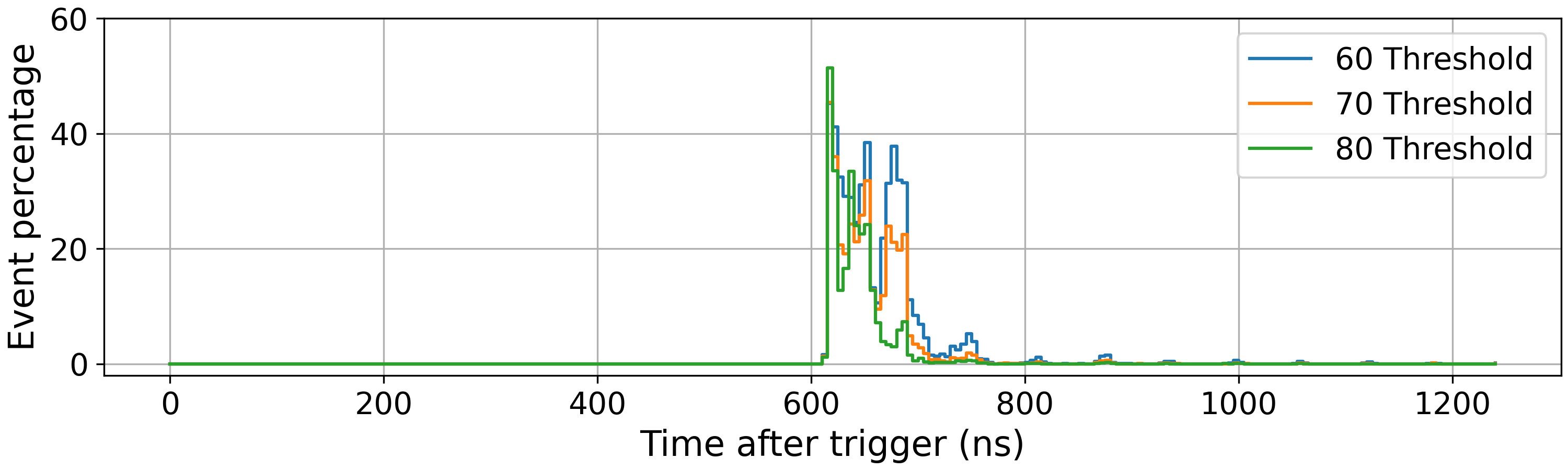}
\caption{Scan for single strip active percentage in a given time bin (5 ns).\label{TimeNoise_30_60}}
\end{figure}

Figure~\ref{TimeNoise_30_60} shows the data recorded within this time frame. Each bin entry indicates the percentage of data recorded in that specific time bin exhibiting a single strip event. Notably, the data recorded at lower threshold values reveals the inherent base noise in the system. As the threshold increases, the baseline noise progressively decreases, eventually reaching low levels at a threshold value of 50. This observation provides valuable insights into the noise characteristics and threshold adjustments for optimal data quality.

\subsection{Filters}
To enhance the detection of real muons while minimizing noise, several filters were developed. These filters are designed based on considerations such as the noise level, time information of the muon following the trigger (depends on RPC time resolution), cluster size of the charge within the detector, and the overall muon flux.

\subsubsection{Timing filter}
\begin{figure}[htbp]
\centering
\includegraphics[width=.82\textwidth, trim=0cm 0 0cm 0cm, clip]{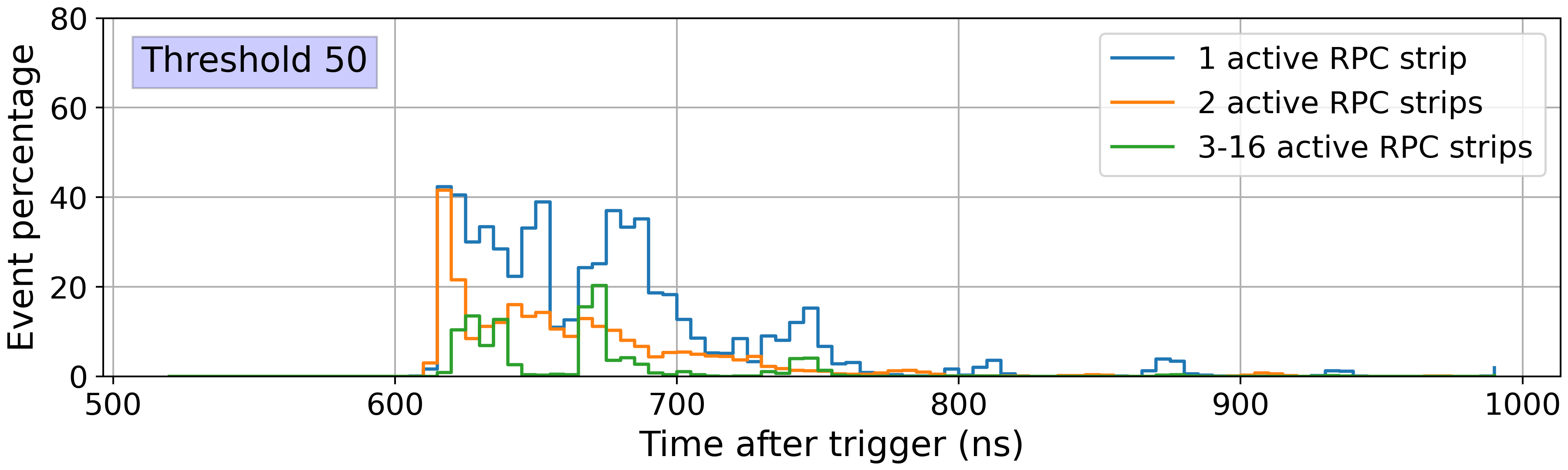}
\includegraphics[width=.82\textwidth, trim=0cm 0 0cm 0cm, clip]{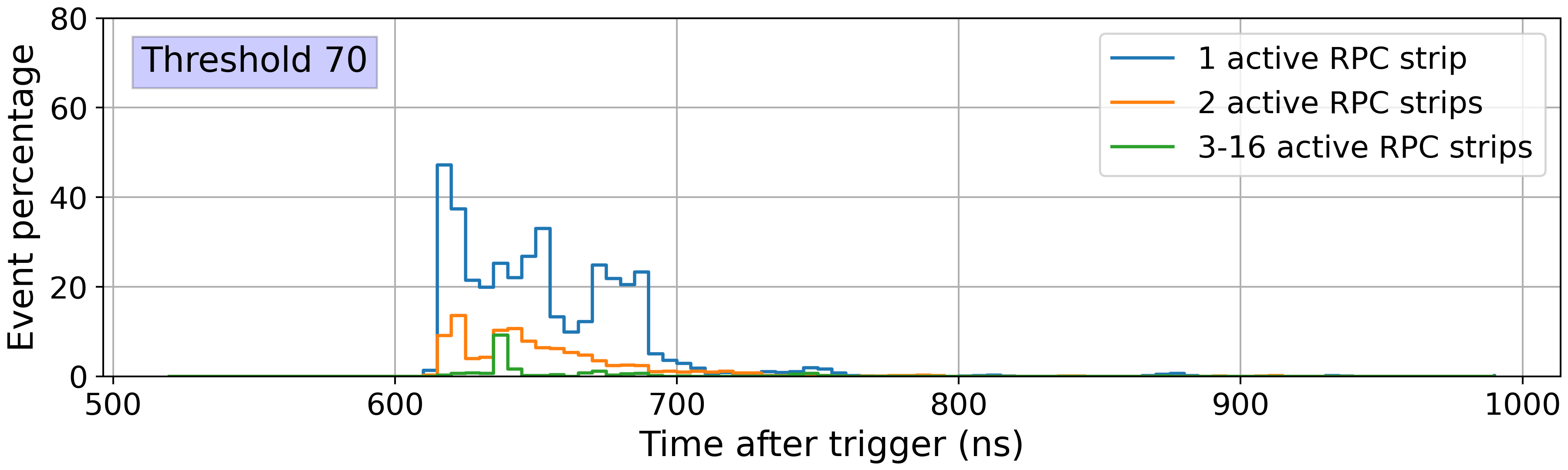}
\includegraphics[width=.82\textwidth, trim=0cm 0 0cm 0cm, clip]{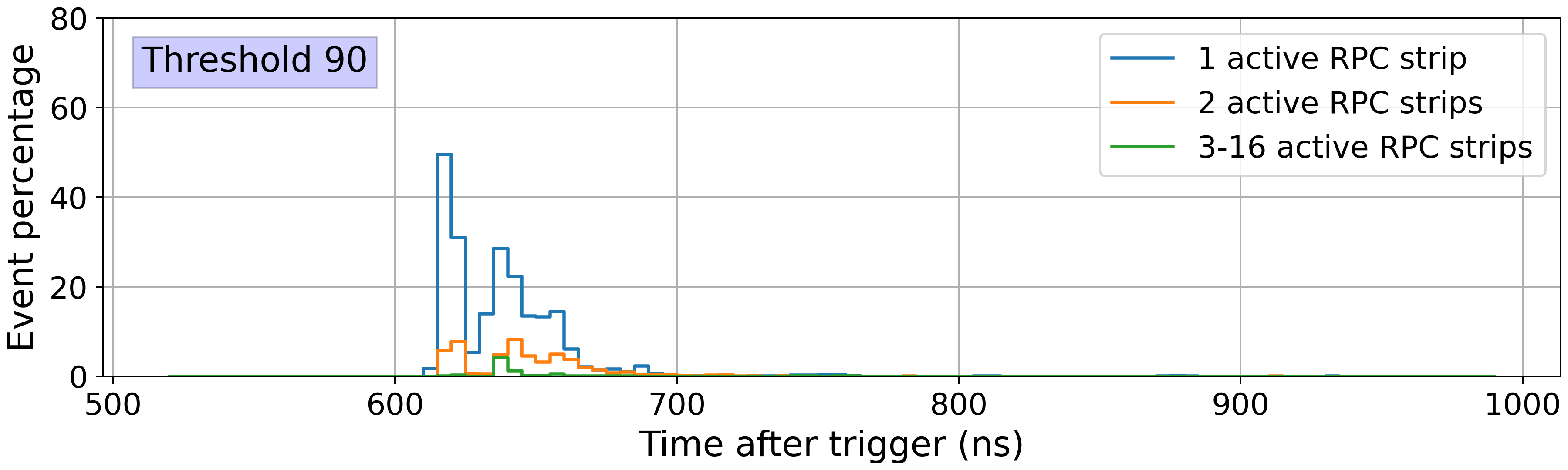}
\caption{Scan for single, double and multiple strip active percentage in a given time bin (5 ns) for 50, 70 and 90 threshold values, respectively.\label{TimeNoise_50_90}}
\end{figure}

Figures \ref{TimeNoise_30_60} and \ref{TimeNoise_50_90} illustrate the impact of threshold variation on signals appearing approximately 620 ns after the trigger. Given that the signal arrival time falls within a range of a few tens of nanoseconds, subsequent signals are attributed to reflections and parasitic capacitance resulting from impedance mismatches~\cite{ammosov2000study, kumar2018effects}. To mitigate the effects of fake coincidence events and noise arising from impedance issues, a timing filter of 15 ns (610-625 ns) is applied. This filtering process enhances the accuracy of the observed events.

\subsubsection{Strip multiplicity filter}
A filter is applied such that events with a strip multiplicity of 1 or 2 are selectively considered. This choice is made based on our understanding of the signal spread due to the charge distribution of the avalanche/streamer and the resistivity of the electrodes. The resistive coating plays a crucial role in localizing the event and ensuring that RPC performance is not affected in other parts of the detector~\cite{thoker2020effect,ye2008studies,das2022studies}. Events with higher multiplicities are likely attributed to noise or the occurrence of a substantial streamer. The positional information of such events is thus deemed less significant for the analysis at hand.
  
\subsubsection{Number of clusters per event}
We define as ``cluster'' a set of adjacent strips, simultaneously giving signals above threshold. Ideally, each cluster corresponds to a single muon event.  
Since our active area is $16 \times 16~{\rm cm}^2$ and the muon flux is very small relative to the timescale in which we are operating (i.e., 15~ns after the time filter), the likelihood of having more than one muon hitting a detector simultaneously is negligible. Therefore, the data filter considers only one cluster per detector per event, under the assumption that in events with two clusters the second one is more likely to originate from noise than from a second muon. 
For illustration, events such as "0011000" (i.e., from left to right, two consecutive strips below threshold followed by two consecutive above threshold and then three consecutive below threshold) or "0001000" are considered acceptable, while "0010010" or "0100110" are not.

\begin{figure}[htbp]
\centering
\includegraphics[width=.32\textwidth, trim=0cm 0 0cm 0cm, clip]{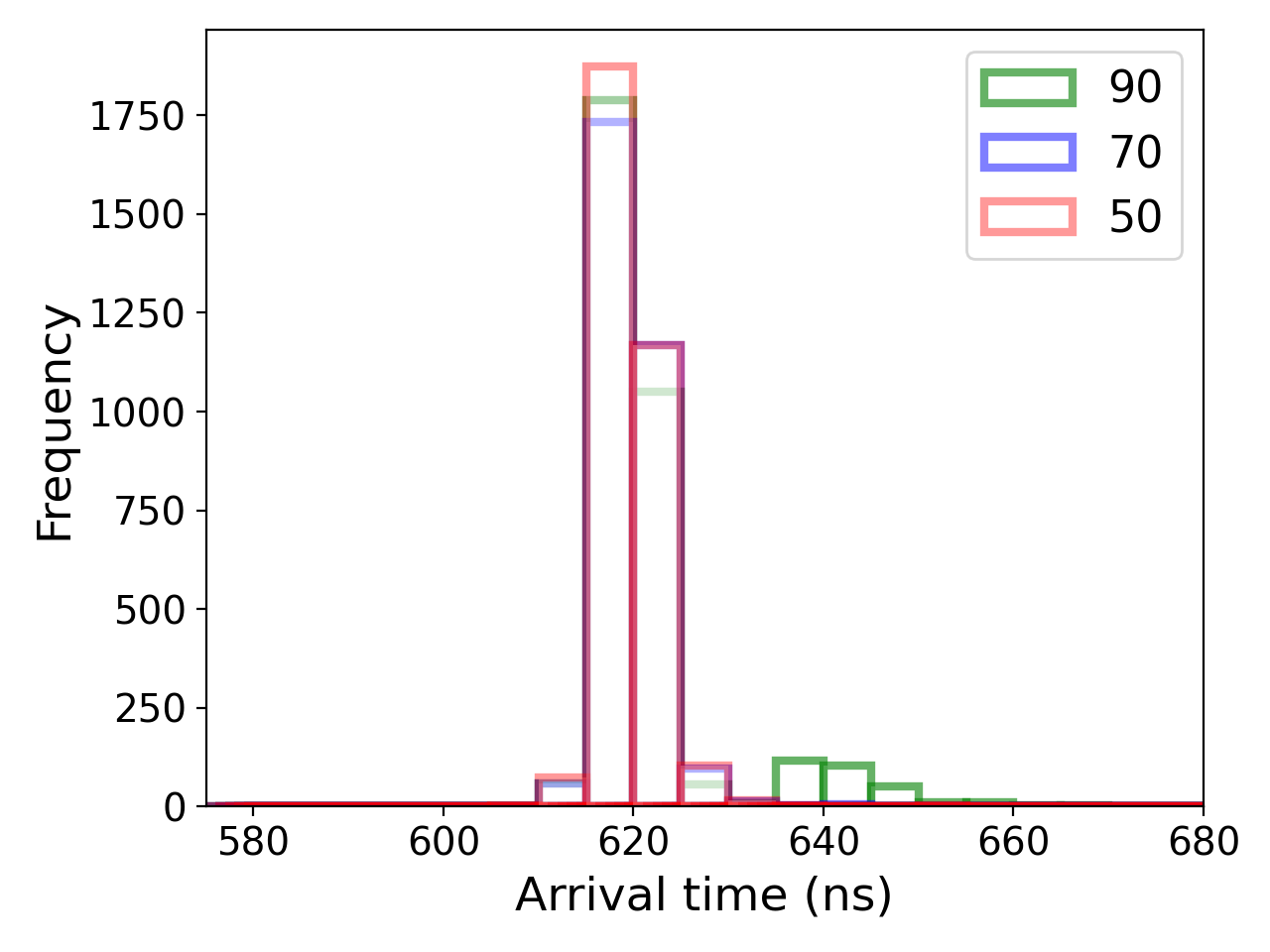}
\includegraphics[width=.32\textwidth, trim=0cm 0 0cm 0cm, clip]{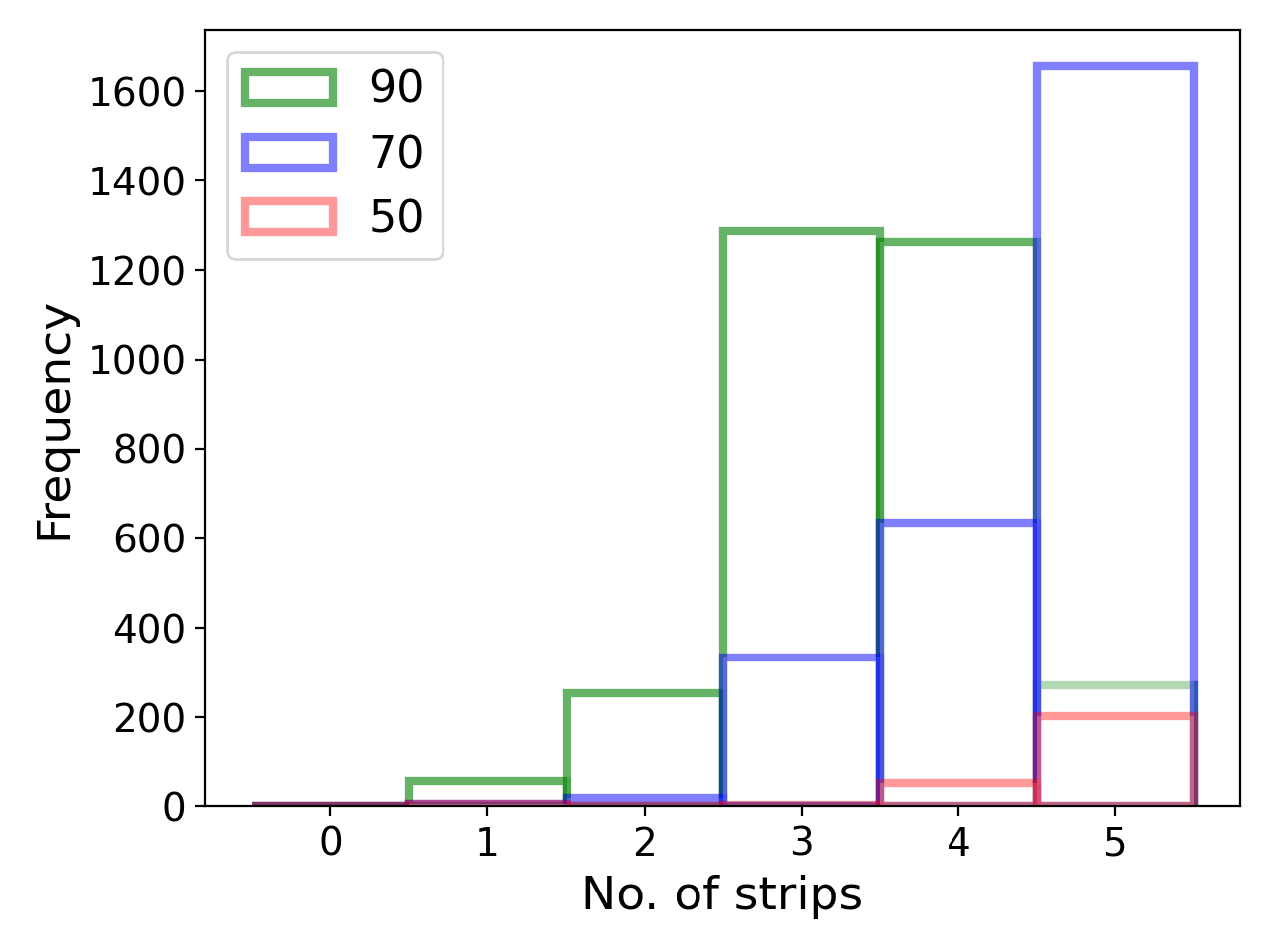}
\includegraphics[width=.32\textwidth, trim=0cm 0 0cm 0cm, clip]{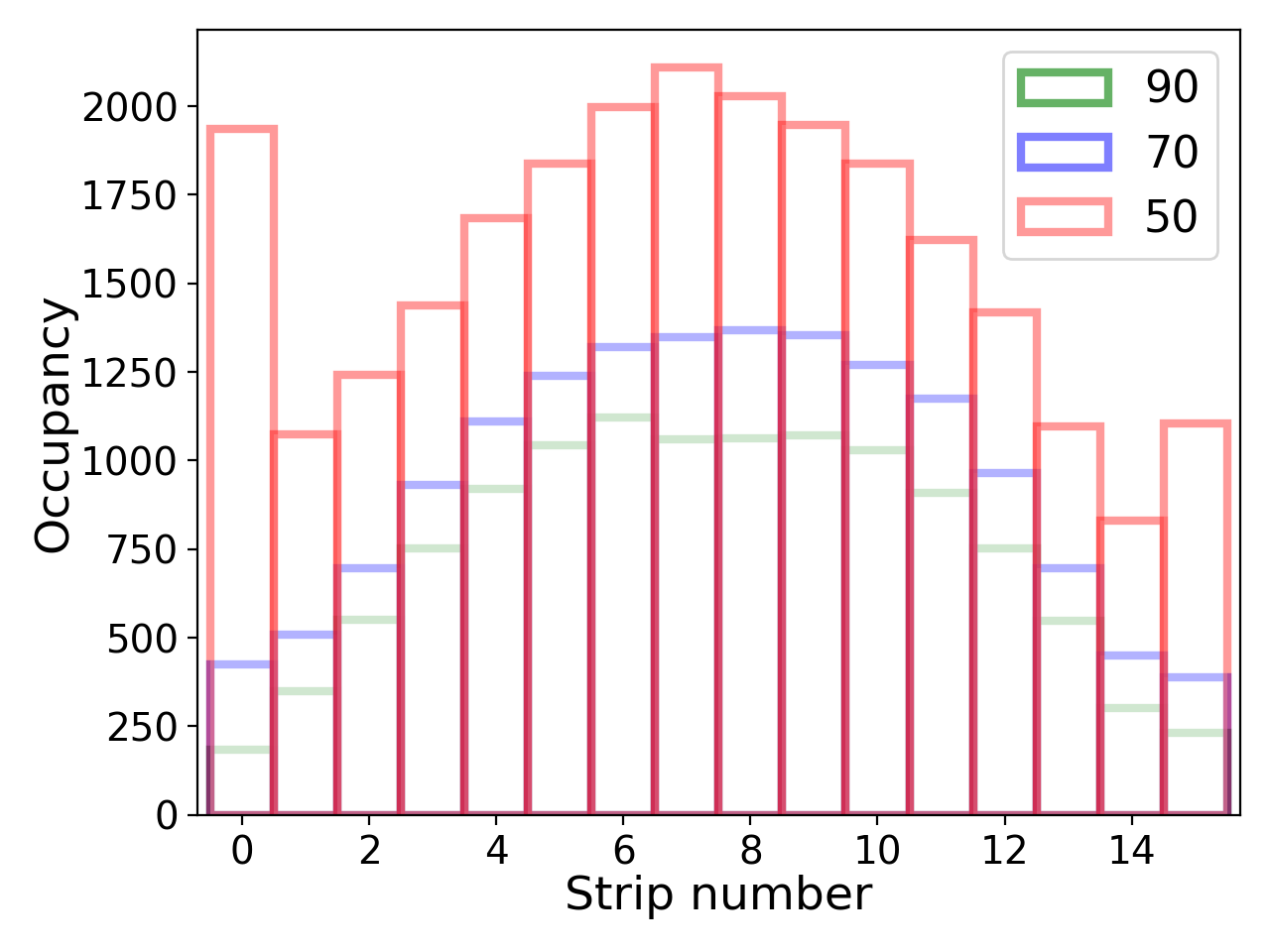}

\includegraphics[width=.32\textwidth, trim=0cm 0 0cm 0cm, clip]{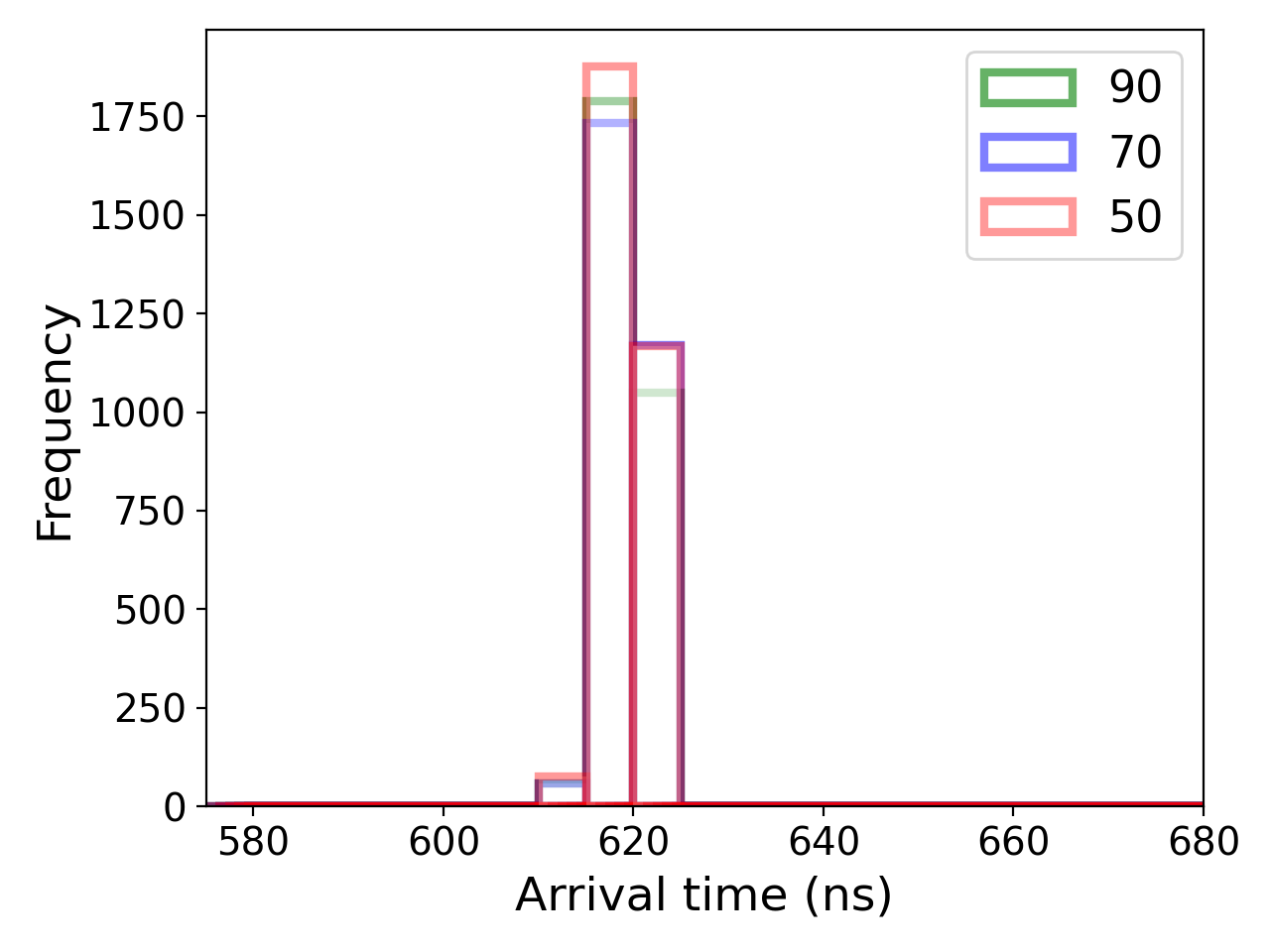}
\includegraphics[width=.32\textwidth, trim=0cm 0 0cm 0cm, clip]{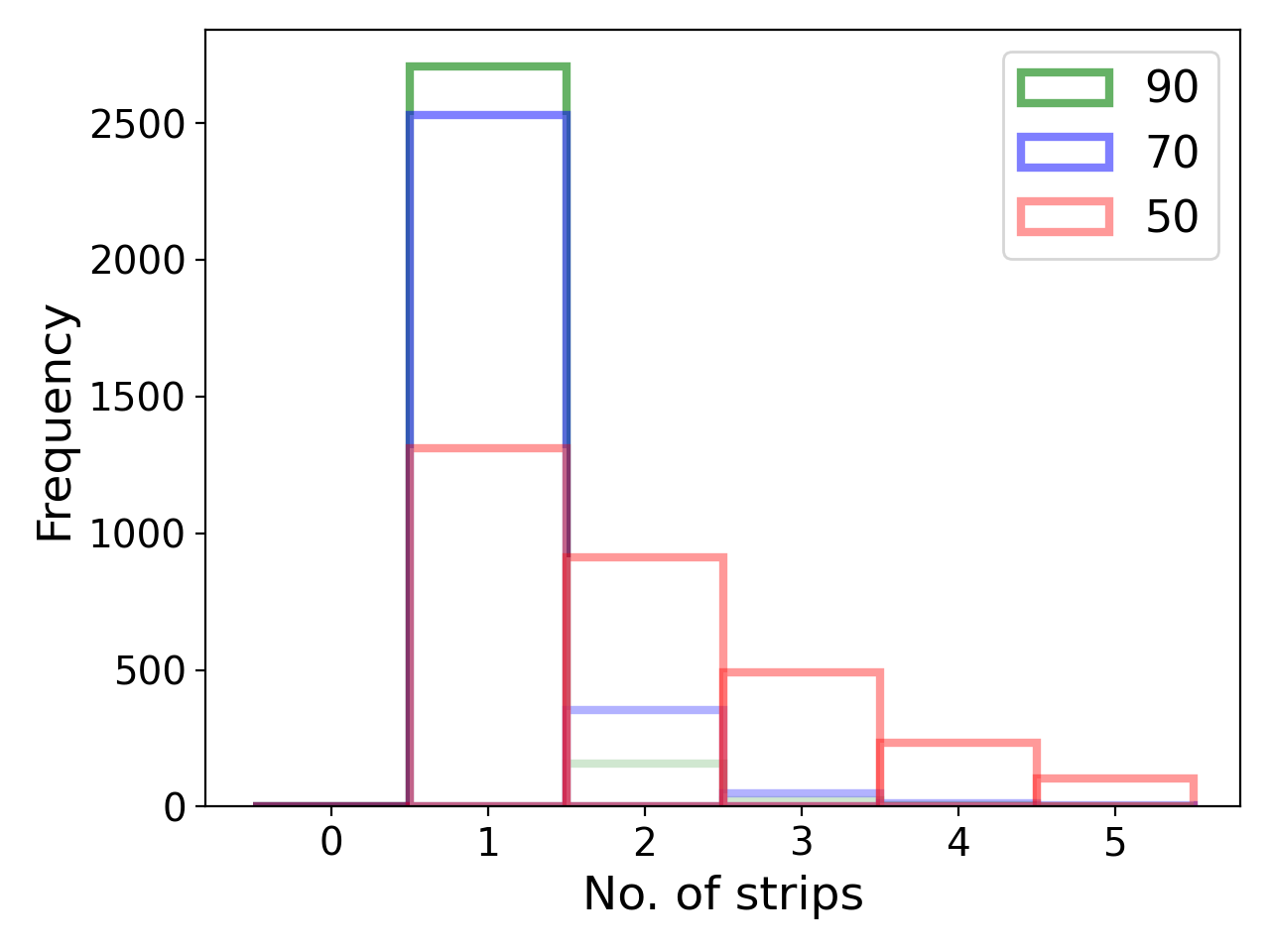}
\includegraphics[width=.32\textwidth, trim=0cm 0 0cm 0cm, clip]{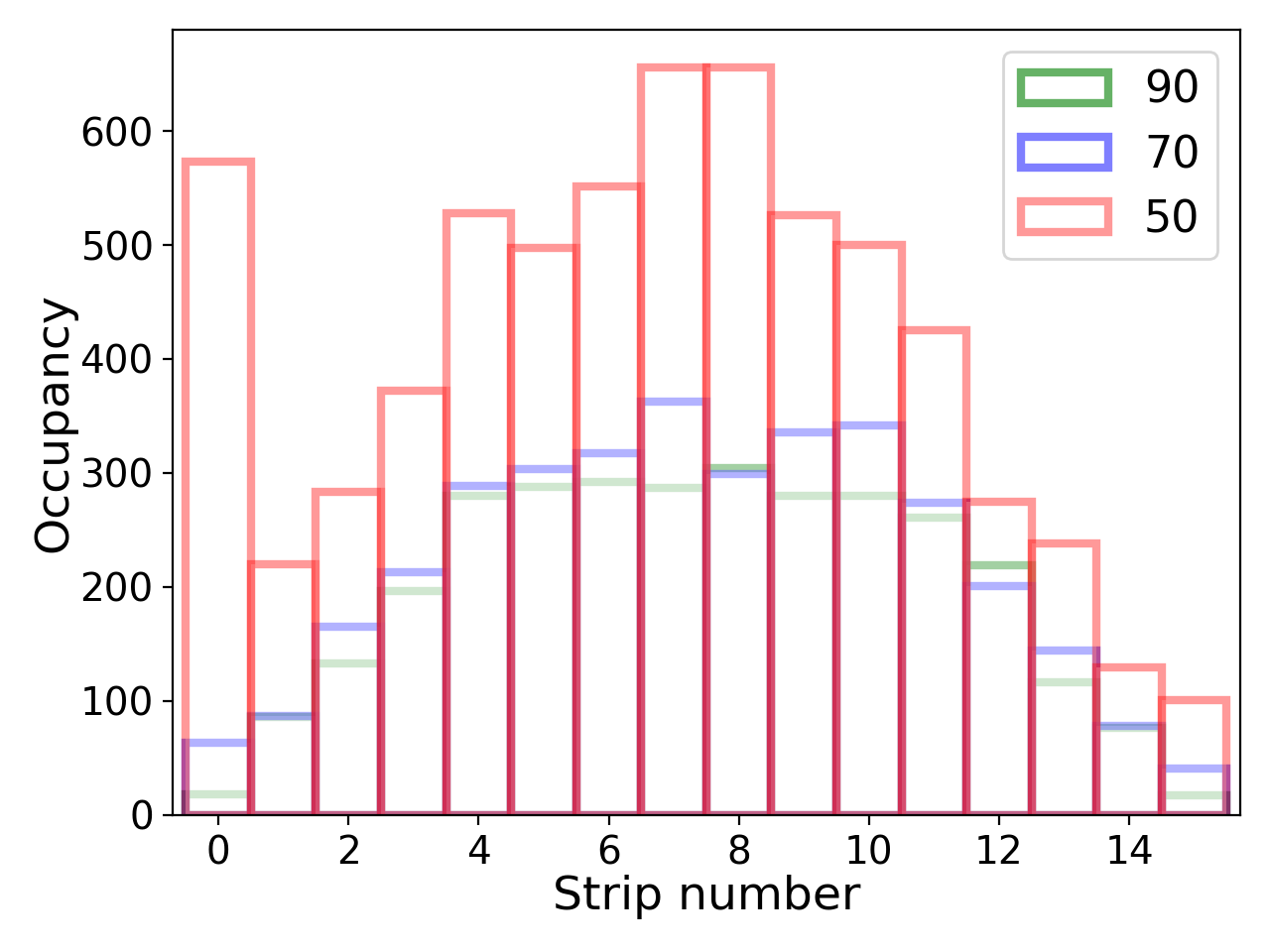}

\includegraphics[width=.32\textwidth, trim=0cm 0 0cm 0cm, clip]{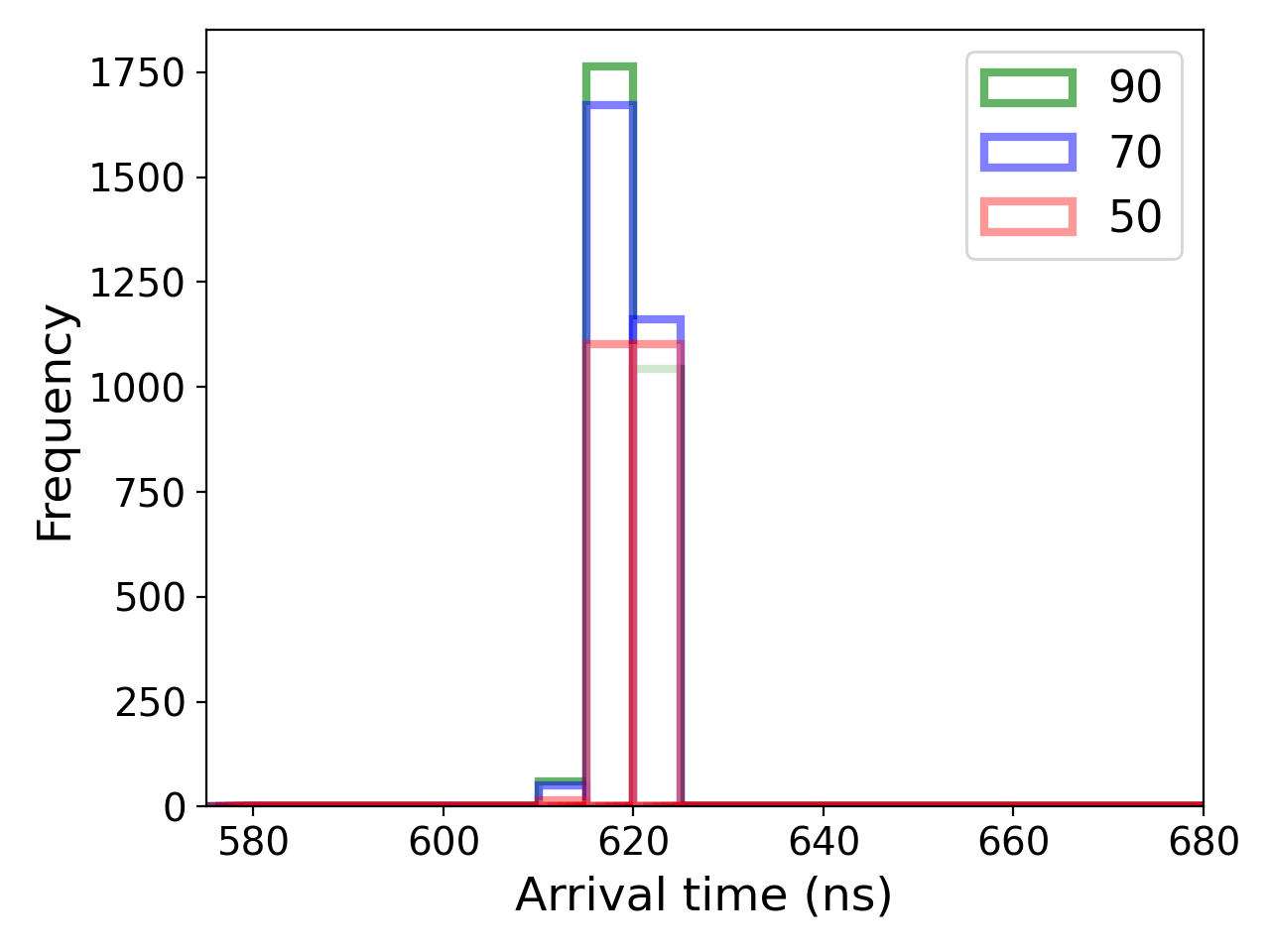}
\includegraphics[width=.32\textwidth, trim=0cm 0 0cm 0cm, clip]{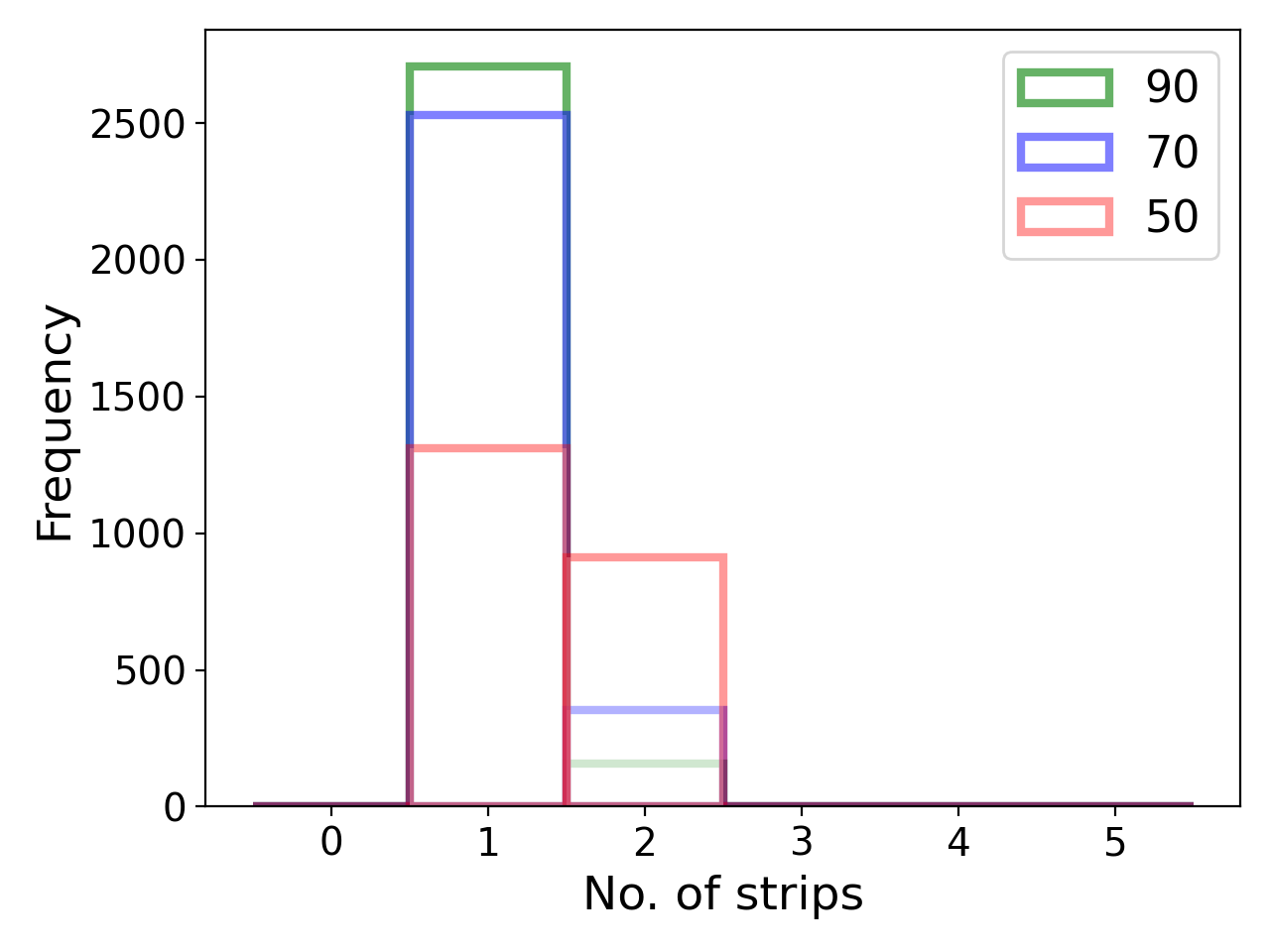}
\includegraphics[width=.32\textwidth, trim=0cm 0 0cm 0cm, clip]{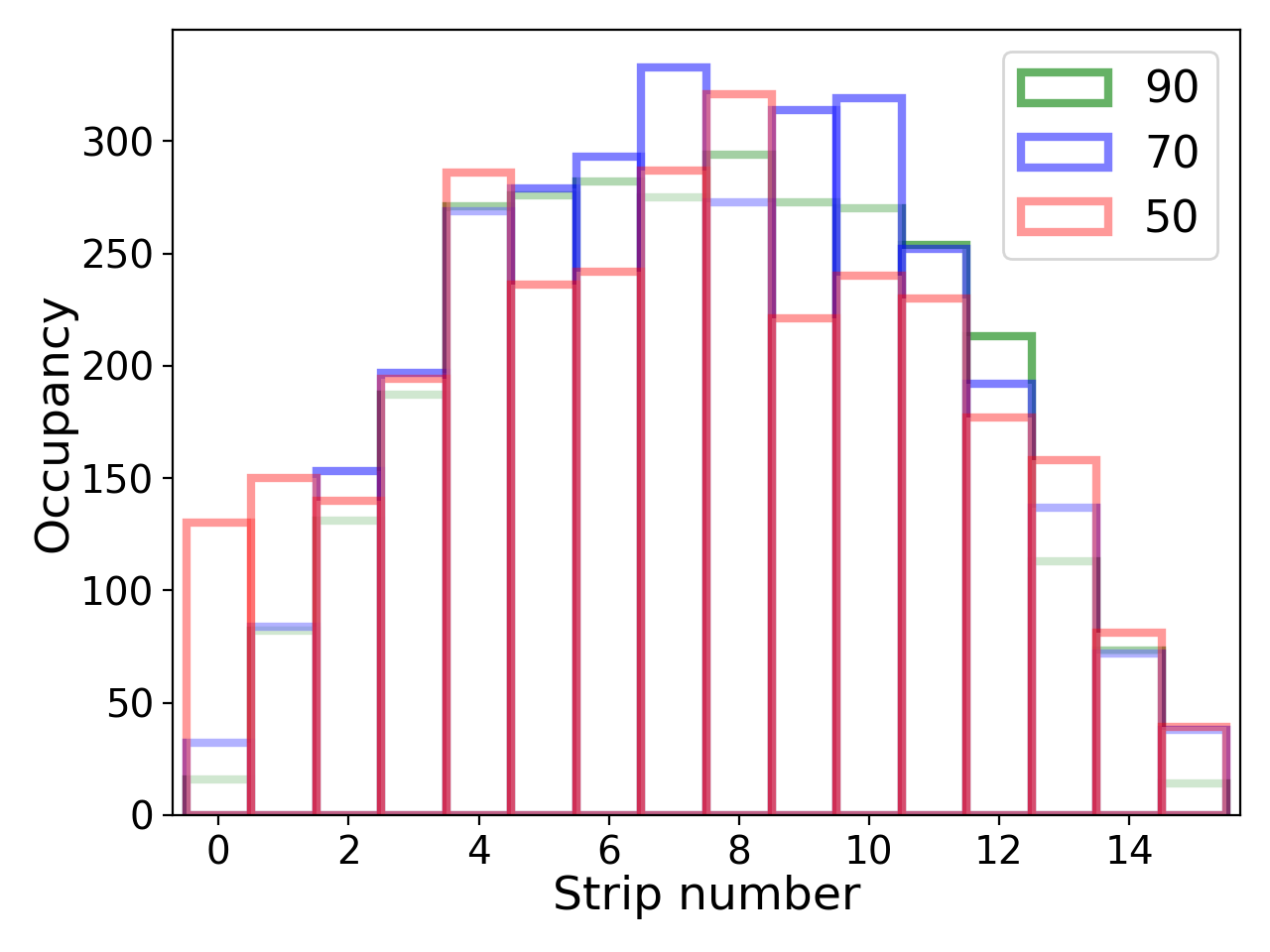}

\includegraphics[width=.32\textwidth, trim=0cm 0 0cm 0cm, clip]{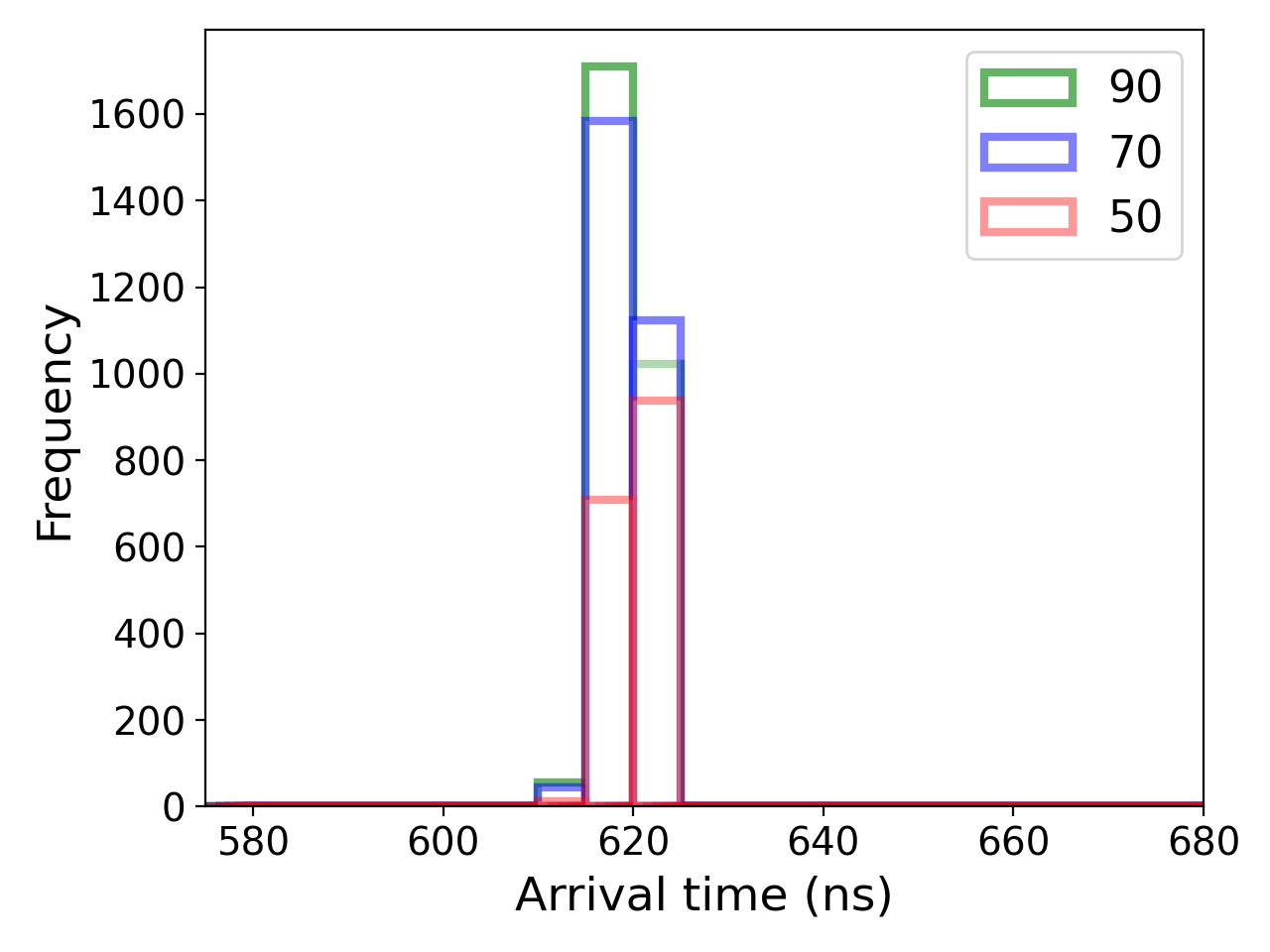}
\includegraphics[width=.32\textwidth, trim=0cm 0 0cm 0cm, clip]{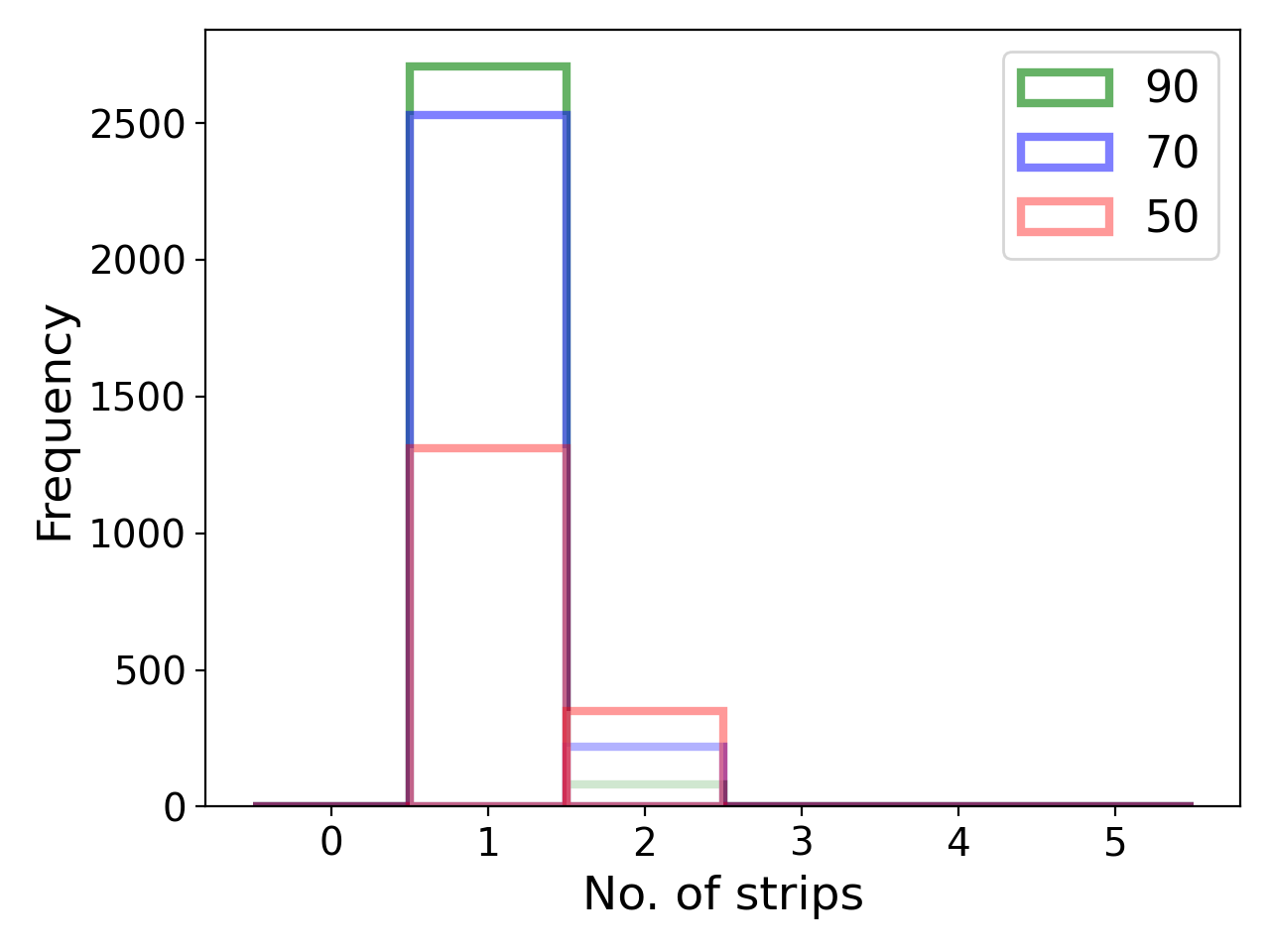}
\includegraphics[width=.32\textwidth, trim=0cm 0 0cm 0cm, clip]{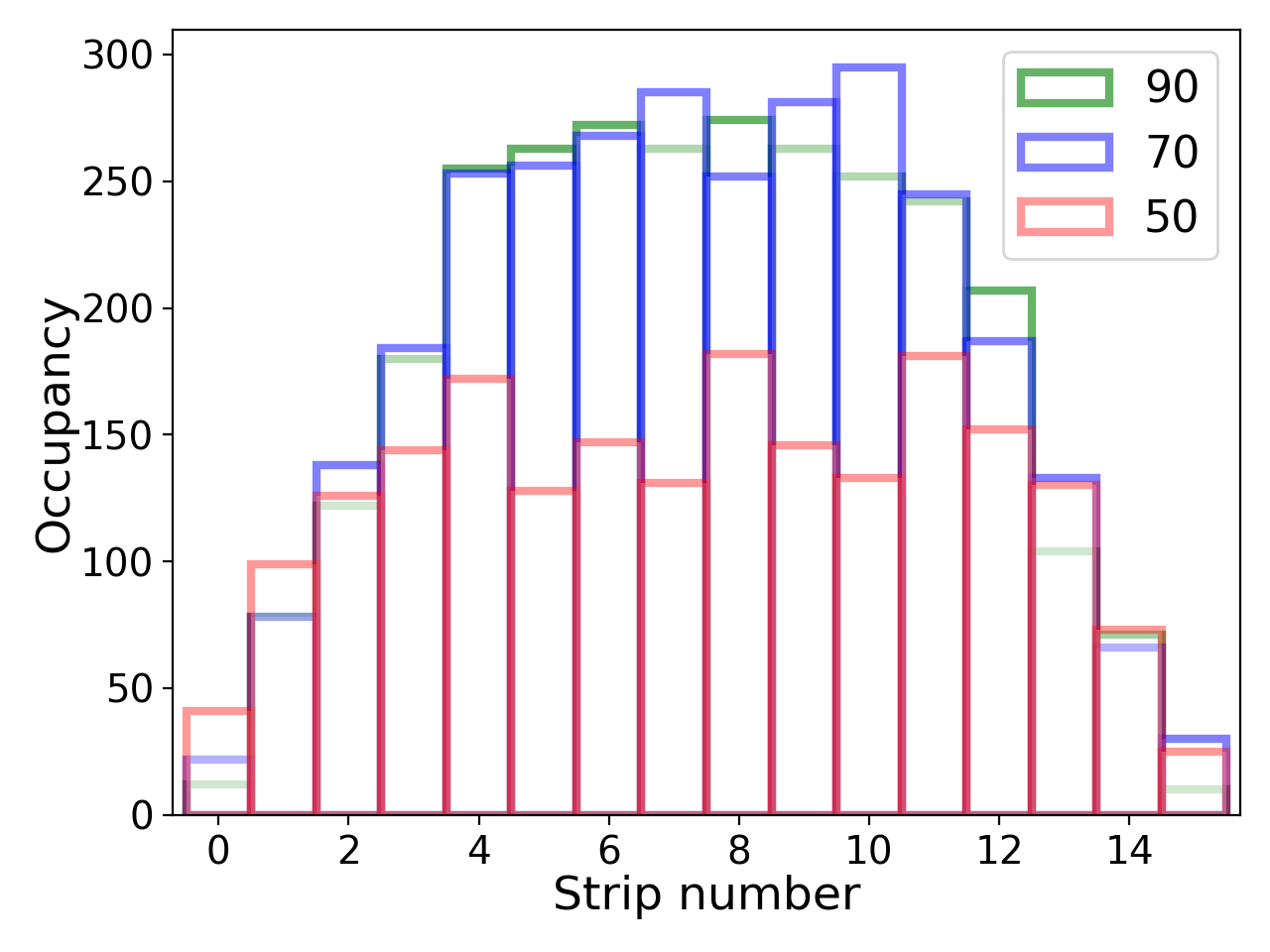}

\caption{
Arrival time information and sequential impact of filters on strip multiplicity and occupancy number, plotted from left to right column-wise. Each row illustrates the cumulative effect of filters: the first row with no filter, the second row with a time filter, the third row with a strip multiplicity filter, and the last row with number of cluster (one in our case) filter. The filters are applied sequentially, with each subsequent filter building upon the results of the preceding ones.\label{filters}}

\end{figure}

Figure~\ref{filters} presents arrival time, strip multiplicity, and strip occupancy for threshold values of 50, 70, and 90, respectively from left to right. The first row displays data with no filters applied, and as evident from figure~\ref{TimeNoise_30_60}, the arrival time for signals is around 620 ns as background noise diminishes at and above 50 threshold. However, the strip multiplicity and occupancy numbers are high due to the cumulative effect of noise resulting from impedance mismatch.

To address this issue offline, a time filter has been implemented, as depicted in the second row. This filter effectively mitigates higher multiplicities, leading to more reasonable strip multiplicity numbers, with a predominant number of events exhibiting only a single active strip. Furthermore, by applying a multiplicity filter to the data, the noise in strip occupancy reduces. The impact is notably more significant for a threshold of 50, resulting in a more reasonable occupancy plot, considering the detector's acceptance due to the geometrical factors of our setup.
Finally, by applying a filter to restrict the number of clusters to one, genuine muon events are notably increased. The impact of these filters on efficiency is presented in table~\ref{table_filters}. These data were taken at a 6.6 kV voltage configuration.

\begin{table}[htbp]
\centering
\caption{Efficiency after the sequential offline filters, for various thresholds. \label{table_filters}}
\smallskip
\begin{tabular}{|c|c|c|c|}
\hline
Threshold & 50 & 70 & 90 \\
No. of events in PMTs & 3479 & 3298 & 3483 \\
\hline
No filter & 93.4 $\pm$ 0.4 \% & 93.5 $\pm$ 0.4 \% & 93.7 $\pm$ 0.4 \% \\
Time filter (610-625 ns) & 89.7 $\pm$ 0.5 \% & 89.7 $\pm$ 0.5 \% & 83.2 $\pm$ 0.6 \% \\
No. of strips $\le$2 & 63.8 $\pm$ 0.8 \% & 87.5 $\pm$ 0.6 \% & 82.3 $\pm$ 0.6 \% \\
No. of clusters = 1 & 47.7 $\pm$ 0.8 \% & 83.4 $\pm$ 0.6 \% & 80.0 $\pm$ 0.7 \% \\
\hline
\end{tabular}
\end{table}

\subsection{Threshold and efficiency scan}
\begin{figure}[htbp]
\centering
\includegraphics[width=.46\textwidth, trim=0cm 0 1.5cm 1.cm, clip]{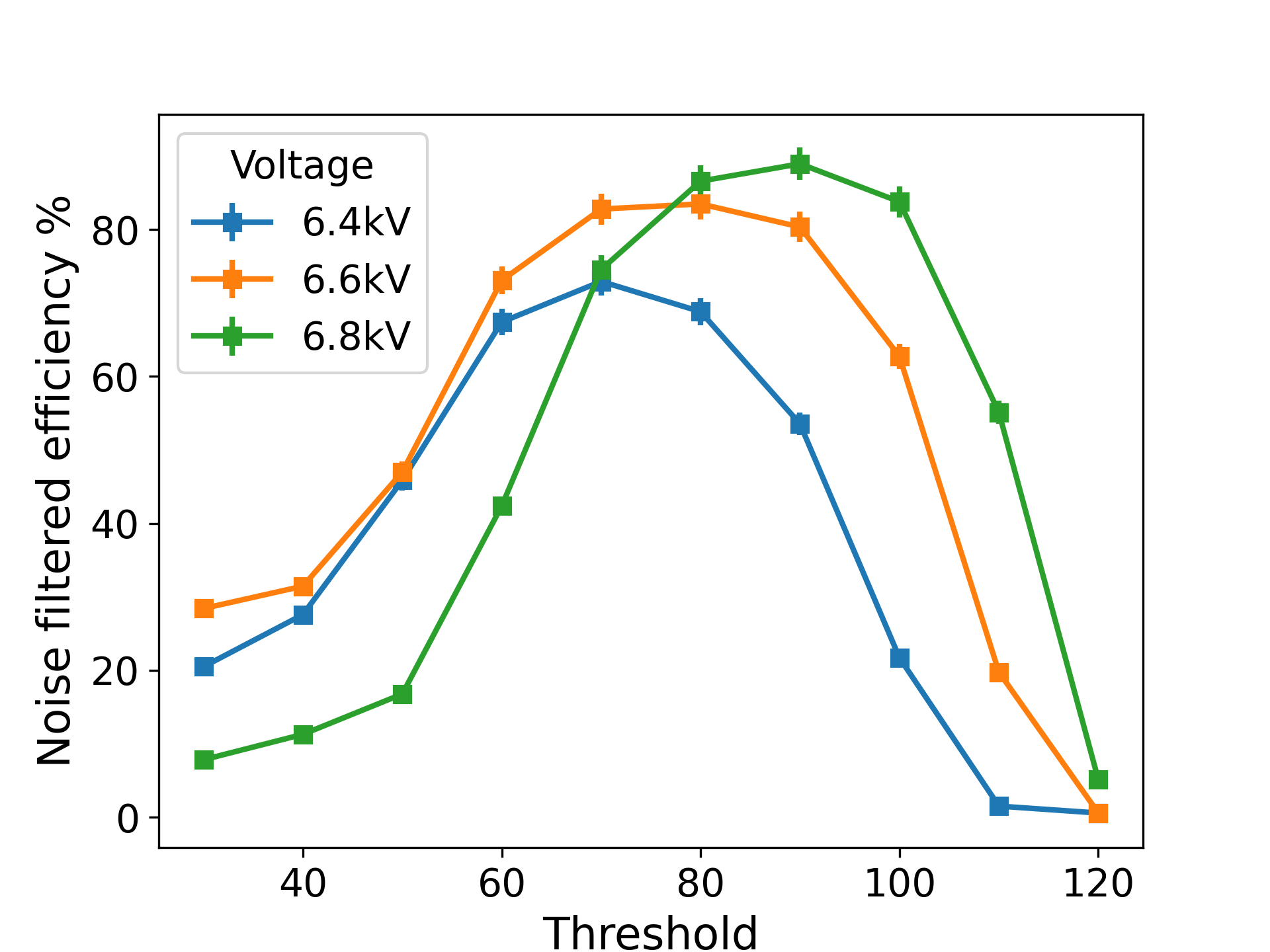}
\caption{Efficiency vs threshold, with all the n    oise filters applied sequentially. Efficiency is lower at low threshold values, due to filtering of real muons along with the noise, and improves as threshold is raised which reduces the noise.\label{Efficiency}}
\end{figure}

To determine the optimal working parameters for the threshold and bias voltage, maximizing the detector's performance, a threshold scan was conducted at various voltage values. These voltage values were thoughtfully selected to ensure a sufficiently high electric field within the RPC for signal formation, either through avalanche or streamer mode. The efficiency versus threshold, used in the front-end board, is plotted in figure \ref{Efficiency} for voltage values of 6.4, 6.6, and 6.8 kV.

Efficiency is calculated by taking the ratio of two-fold coincidence events from the plastic scintillators' Photo-Multiplier Tubes (PMTs) to the number of events in the RPC after applying all the filters mentioned in the previous section. The plot indicates that the efficiency at a low threshold is higher for 6.4 and 6.6 kV compared to 6.8 kV. This discrepancy is attributed to the increased noise level associated with the high-field configuration of the detector. Additionally, it elucidates why the efficiency peaks at a relatively higher threshold with an increase in bias voltage, as the applied filters effectively reduce the signal containing noise.
The maximum efficiency, approximately 88$\%$, is achieved for 6.8 kV at a threshold value of 90.

It is crucial to emphasize that efficiency is compromised at lower threshold values, primarily due to the filtration of genuine muons being intertwined with the noise. However, this efficiency progressively enhances with an increase in the threshold, effectively mitigating the impact of noise. Upon reaching an optimal threshold value,  efficiency peaks before gradually diminishing as the threshold becomes excessively high, leading to the exclusion of low-amplitude signals associated with muons.

\section{Absorption Muography}
\label{sec:Absorption}
\begin{figure}[htbp]
\centering
\includegraphics[width=.5\textwidth, trim=0.3cm 0cm 0.5cm 0.5cm, clip]{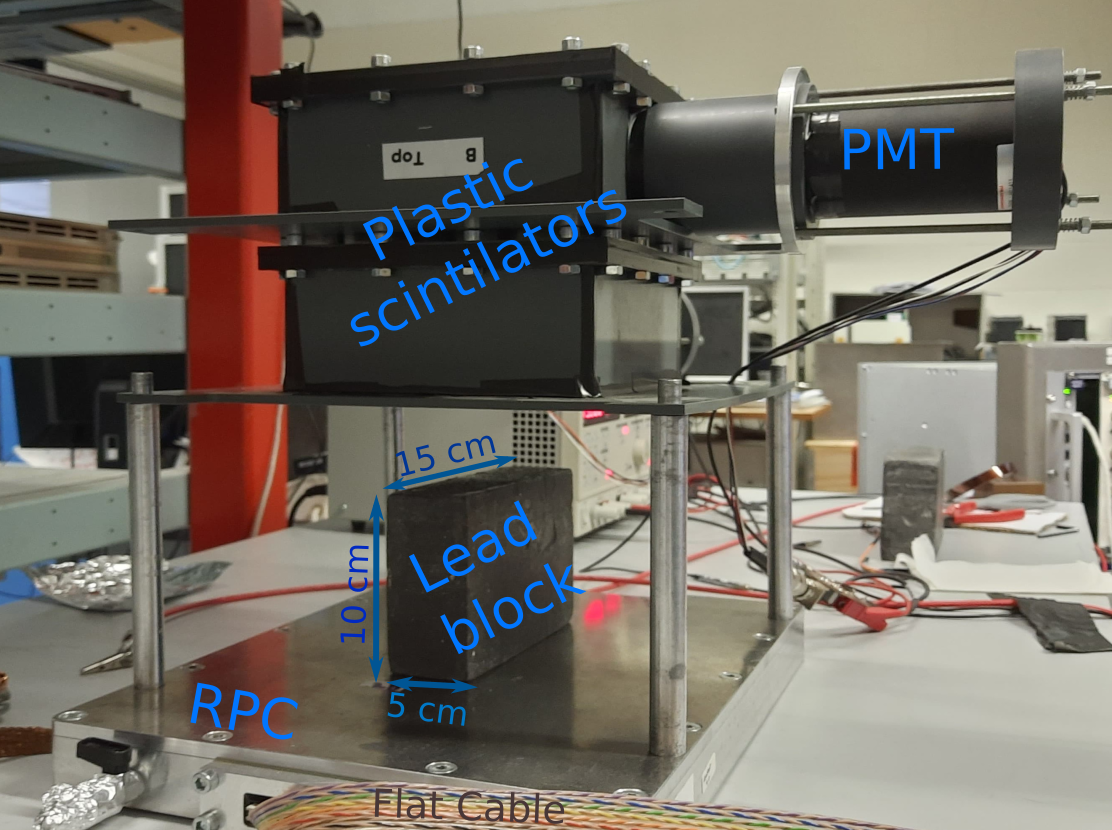}
\caption{Experimental setup for muon absorption tomography.\label{Absorption_setup}}
\end{figure}

To further evaluate the single-detector performance, along with the DAQ and analysis methodologies, a simplified muon absorption experiment has been conducted. The experimental setup is depicted in figure~\ref{Absorption_setup}, featuring two plastic scintillators in conjunction with PMTs positioned 15 cm above the RPC chamber and supported by four cylindrical aluminum pillars. These pillars serve the dual purpose of offering structural support and defining a scanning volume, commonly referred to as the Region Of Interest (ROI).
The signals obtained from the PMTs serve as trigger signals for data collection from the RPC, similar to the process used in efficiency measurement. In this experiment, two sets of data were recorded: one with a lead block placed in the ROI and another without it (blank), providing a baseline for comparison. The lead block has dimensions of $15 \times 5 \times 10~{\rm cm}^3$, with a length of 15 cm parallel to the RPC's readout strips, a width of 5 cm perpendicular to the strips, and a height of 10 cm. The lead block is positioned close to the center, covering approximately strip numbers 5 to 9.

\begin{figure}[htbp]
\centering
\includegraphics[width=.45\textwidth, trim=0cm 0 .9cm .5cm, clip]{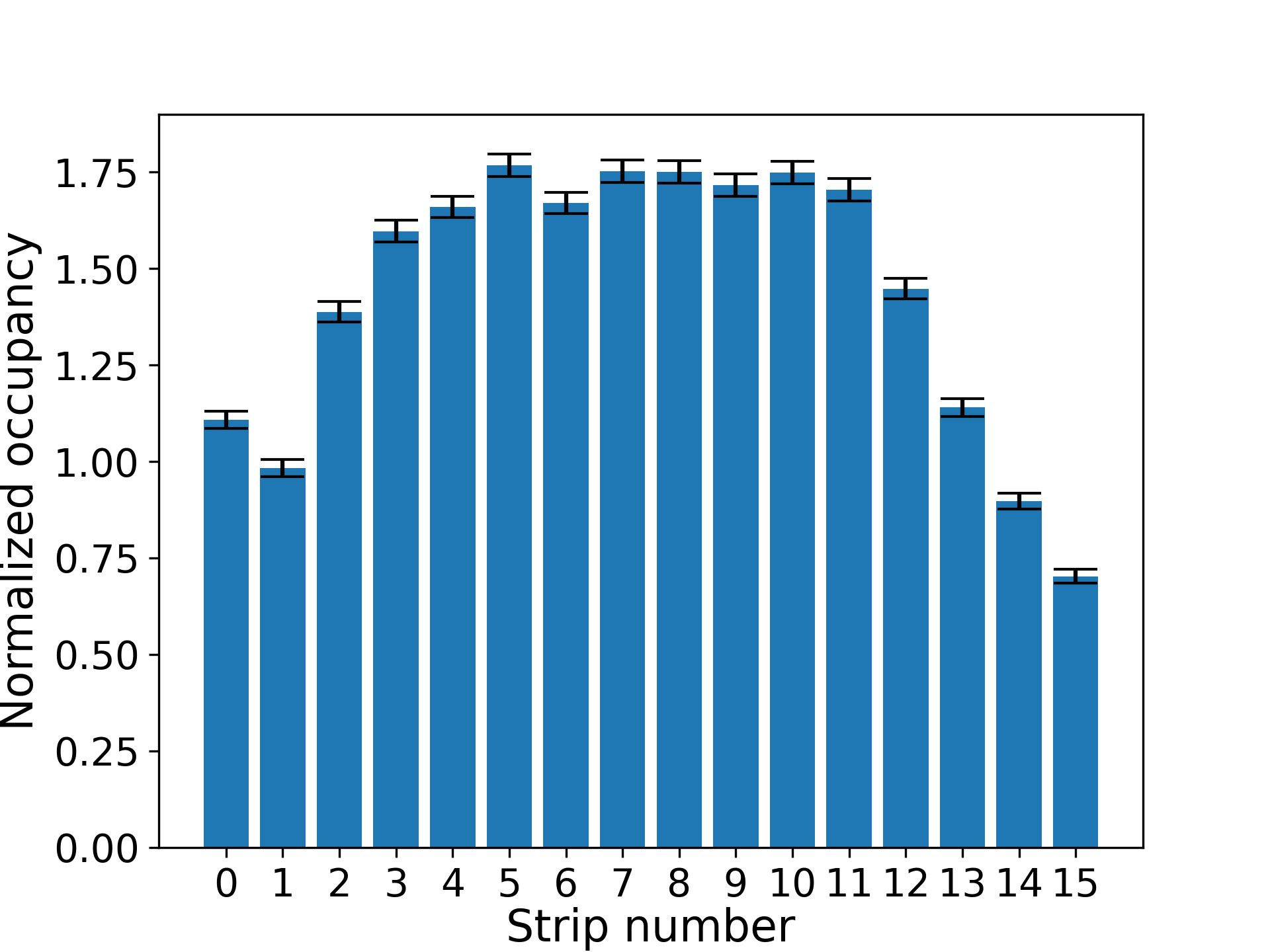}
\includegraphics[width=.45\textwidth, trim=0cm 0 .9cm .5cm, clip]{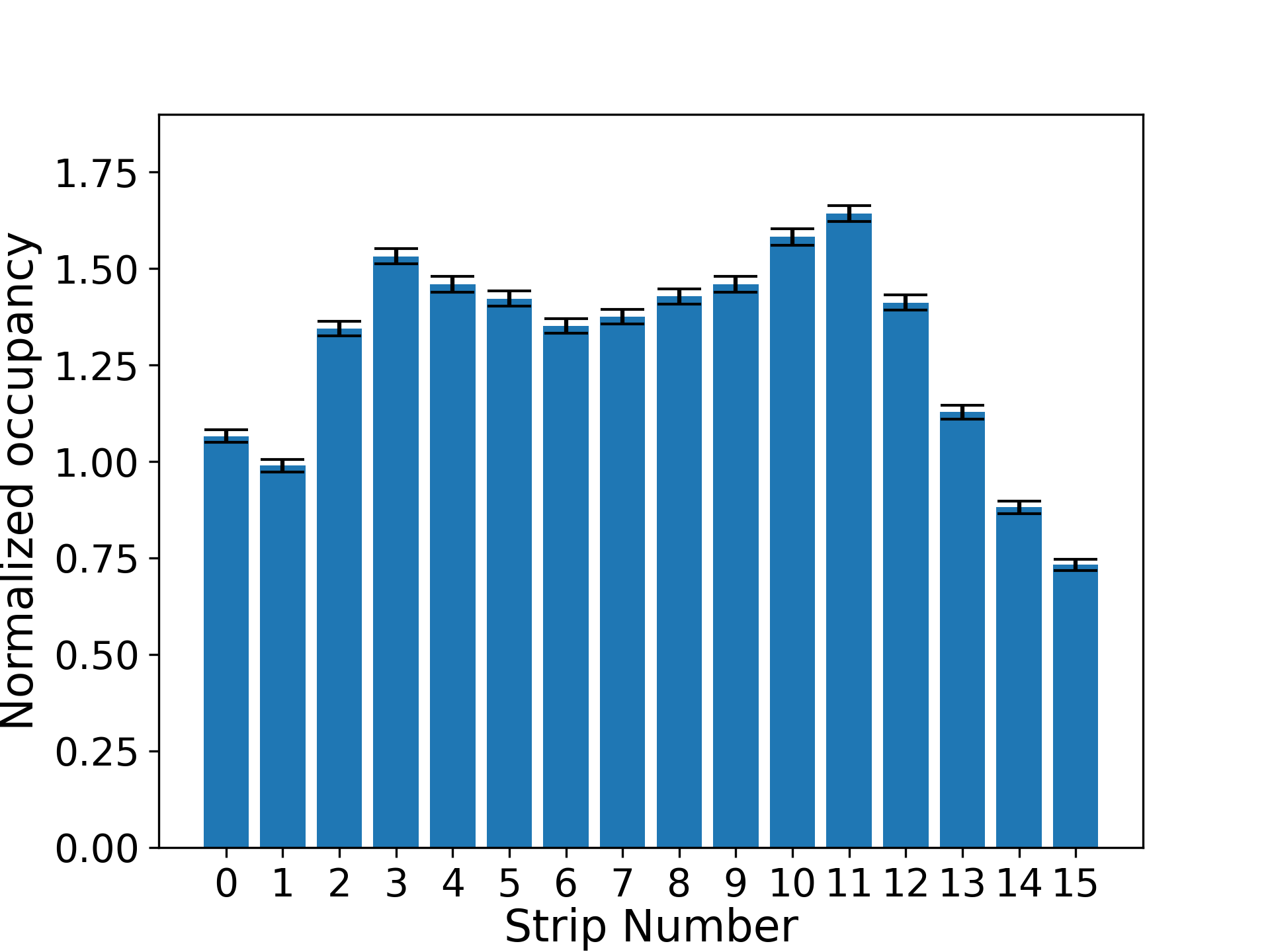}
\caption{Strip occupancy without (on the left) and with (on the right) a lead block placed in the ROI.\label{Absorption_Ocupancy}}
\end{figure}

Figure~\ref{Absorption_Ocupancy} displays normalized occupancy numbers for two scenarios: one with blank data (left plot) and the other with the lead block (right plot). Normalized occupancy is calculated as the percentage of events in a strip relative to the total triggers received by the PMTs. This normalization is crucial for mitigating statistical fluctuations in muon flux throughout the data collection period. Examining the normalized occupancy plot reveals a noticeable reduction of muon counts in correspondence of the lead block position and its surroundings.

\begin{figure}[htbp]
\centering
\includegraphics[width=.45\textwidth, trim=0cm 0 .9cm .5cm, clip]{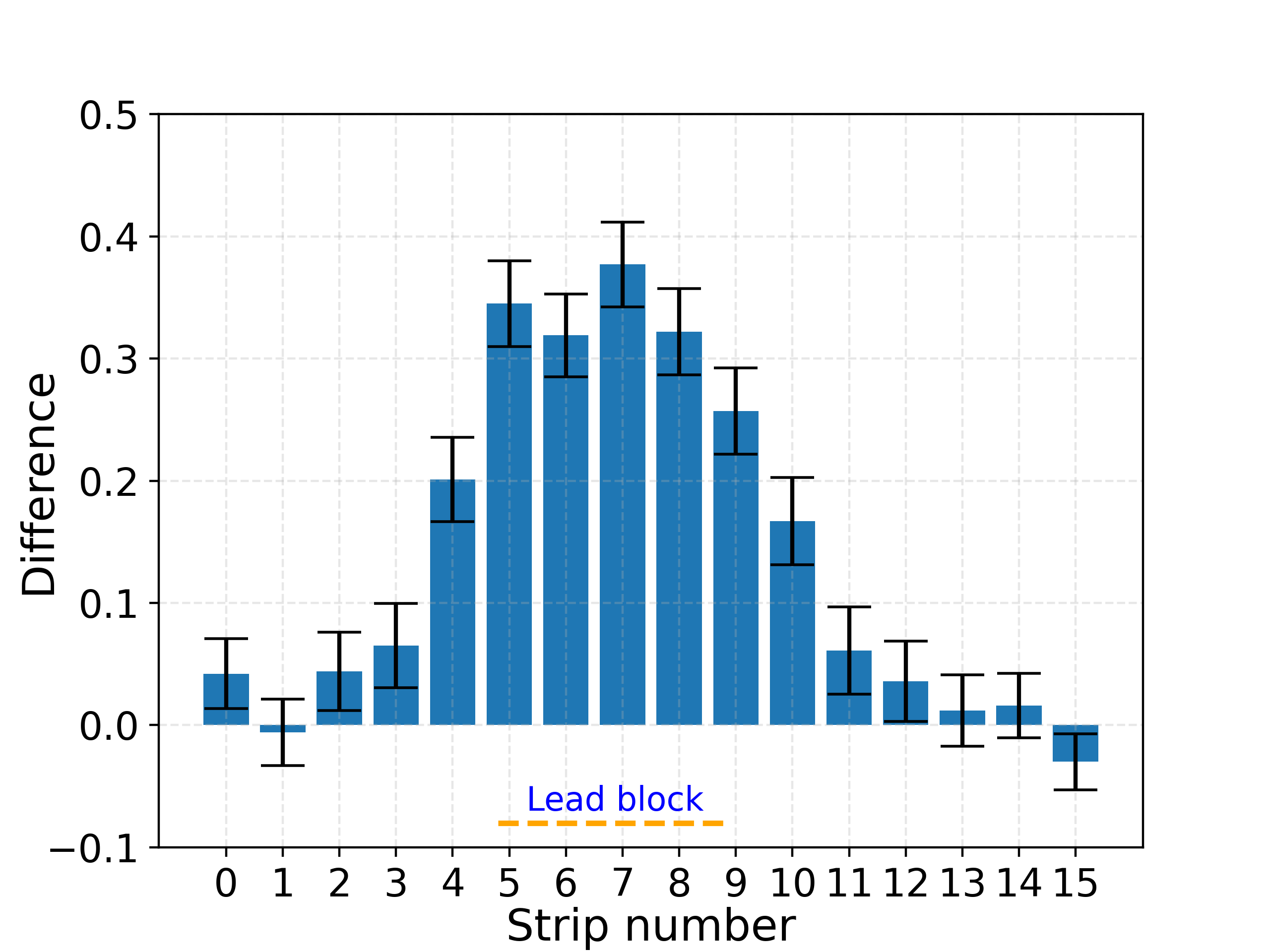}
\includegraphics[width=.45\textwidth, trim=0cm 0 .9cm .5cm, clip]{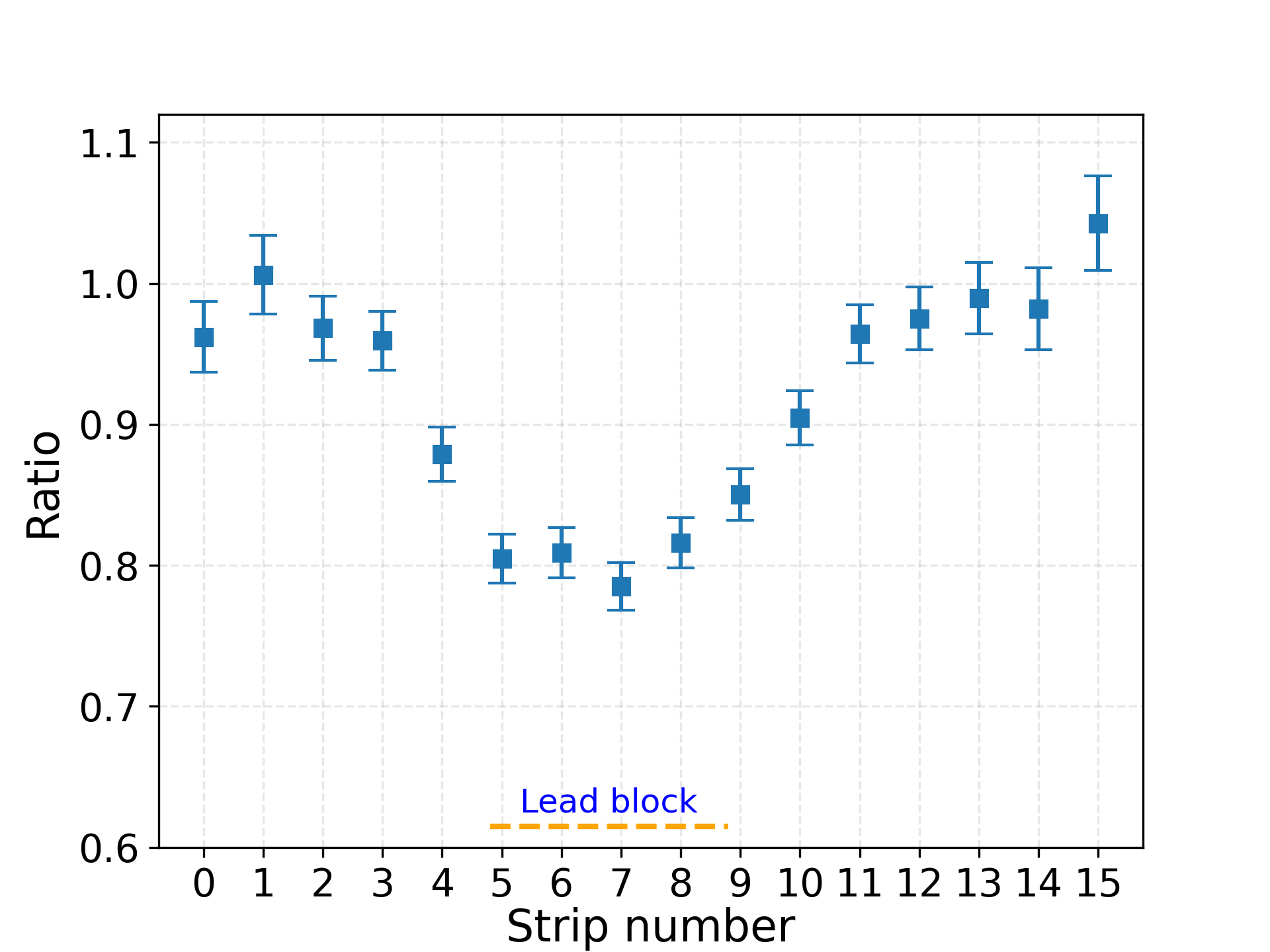}
\caption{Difference (on the left) and ratio (on the right) of the normalized occupancy value after the introduction of a lead block in the ROI with respect to the blank.\label{Absorption_Study}}
\end{figure}

Furthermore, the reduction of muon flux due to absorption is studied by analyzing the difference and ratio of normalized strip occupancy, as shown in figure~\ref{Absorption_Study}. The difference plot depicts the strip-wise variance in the number of muons before placing the lead block (blank) and after the placement of the lead block. Similarly, the ratio plot shows the strip-wise ratio of the normalized occupancy number before and after the placement of the lead block. 
The plots clearly indicate reduced strip occupancy where the lead block is placed, and the reduction is also observed around the lead block, as some of the muons approaching at an angle from the sides are either absorbed or scattered away from that region. The orange line in the plots shows the actual position of the lead block in the axis perpendicular to the strips.

\section{Summary and Outlook}
\label{sec:conclusions}



A comprehensive performance test of our gas-tight RPC has been successfully conducted, employing a meticulous pulse processing and data acquisition methodology. The discussion delves into the intricacies of filtering genuine events from noise, addressing challenges posed by reflections and cross-talk. The impact of these filters on both efficiency and event quality is detailed. Additionally, a detailed threshold scan, exploring various voltage configurations, has been undertaken to determine the optimal performance settings. Finally, a simplified (single-RPC) muon absorption experiment has been executed, providing valuable insights into the feasibility and performance of these gas-tight RPCs from a practical application standpoint. 

Our prototype RPC detectors are currently undergoing long-term operational testing to ensure gas stability, an important factor directly impacting RPC performance. These tests are critical to ensure the performance of the detector in the field environment, and will be reported in a future publication. The testing, conducted on a prototype, will advance the construction of a hodoscope featuring multiple RPCs, each equipped with both x- and y-oriented planes. 
It is anticipated that scintillators may be eliminated in the final setup, as RPCs are designed to be self-triggering. With sufficiently low noise, we can rely on the redundancy provided by having multiple detectors at the trigger level.

\acknowledgments

This work was partially supported by the EU Horizon 2020 Research and Innovation Programme under the Marie Sklodowska-Curie Grant Agreement No. 822185, and by the Fonds de la Recherche Scientifique - FNRS under Grants No. T.0099.19 and J.0070.21. 





\bibliographystyle{unsrt}
\bibliography{biblio}

\end{document}